\documentclass[preprint,groupedaddress,showpacs,showkeys,amsmath]{revtex4}

\usepackage{amssymb}

\newcommand{\kln}{ \ln_{\{ \kappa \}} }
\newcommand{\kexp}{ \exp_{ \{\kappa \}} }
\newcommand{\kprod}{ \mathop{\otimes}_{\kappa} }
\newcommand{\uprod}{ \mathop{\odot}_{\kappa} }
\newcommand{\kdiv}{ \mathop{\oslash}_{\kappa} }
\newcommand{\Odiv}{ {\ooalign{ $\mathop{\ominus}$\crcr{$\mathop{\div}$} }} }
\newcommand{\udiv}{ \mathop{\Odiv}_{\kappa} }
\newcommand{\ku}{ u_{\{ \kappa \}} }

\begin{document}
\title{$\kappa$-generalization of Stirling approximation
and multinominal coefficients}
\author{Tatsuaki Wada} 
\email{wada@ee.ibaraki.ac.jp}
\affiliation{Department of Electrical and Electronic Engineering,
Ibaraki University, Hitachi,~Ibaraki, 316-8511, Japan}

\author{Hiroki Suyari}
\email{suyari@faculty.chiba-u.jp, suyari@ieee.org}
\affiliation{Department of Information and Image Sciences,
Chiba University, Chiba, 263-8522, Japan}

\date{\today}

\begin{abstract}
  Stirling approximation of the factorials and multinominal coefficients are
generalized based on the one-parameter ($\kappa$) deformed functions
introduced by Kaniadakis [Phys. Rev. E \textbf{66} (2002) 056125].
We have obtained the relation between the $\kappa$-generalized 
multinominal coefficients and the $\kappa$-entropy by introducing 
a new $\kappa$-product operation. 
\end{abstract}


\maketitle

\section{Introduction}
As is well-known in the fields of statistical mechanics, Ludwig Boltzmann
clarified the concept of entropy. He considered a macroscopic system
consists of a large number of particles. 
Each particle is assumed to be in one of the energy level 
$E_i \; (i=1,\ldots, k)$
and the number of particles in the level $E_i$ is $n_i$. 
The total number
of particles is then $n = \sum_{i=1}^k n_i$ and the total energy 
of a macrostate is $E=\sum_i n_i E_i$. The number of ways of
arranging $n$ particles in $k$ energy levels such that each
level has $n_1, n_2, \ldots, n_k$ particles are the multinominal coefficients
\begin{equation}
   \begin{bmatrix} n \\ n_{1} \cdots  n_{k}
   \end{bmatrix}
   \equiv \frac{n!}{ n_{1}! \cdots  n_{k}!},
   \label{multi}
\end{equation}
which is proportional to the probability of the macrostates
if all microscopic configurations are assumed to be equally likely.
In the thermodynamic limit $n$ increasing to infinity, we consider
the relative number $p_i \equiv n_i / n$ as the probability of
the particle occupation of a certain energy level $E_i$.
In order to find the most probable macrostate, Eq. \eqref{multi}
is maximized under a certain constraint. The Stirling approximation
of the factorials leads to 
  \begin{equation}
   \ln \begin{bmatrix} n \\ n_{1} \cdots  n_{k}
   \end{bmatrix}
   \approx  n \;
   S^{\rm BGS} \left(\frac{n_1}{n}, \ldots, \frac{n_k}{n}\right),
 \label{entropy}
\end{equation}
where
\begin{equation}
   S^{\rm BGS} \left(\frac{n_1}{n}, \ldots, \frac{n_k}{n}\right)
  \equiv -\sum_{i=1}^k p_i \ln (p_i), 
 \label{BGS}
\end{equation}
is the Boltzmann-Gibbs-Shannon (BGS) entropy.

Quite recently one of the authors (H.S.) has shown one-parameter
generalizations of Gauss' law of error \cite{Suyari05}, Stirling
approximation \cite{Suyari-a}, and multinominal coefficients
\cite{Suyari-b} based on the Tsallis $q$-deformed functions and
associated multiplication operation ($q$-product) \cite{Wang,Borges}.
These mathematical structures are quite fundamental for the basis
of any generalization of statistical physics as in the standard
statistical physics. 
In particular Ref. \cite{Suyari-b} has shown that the one-to-one 
correspondence between the $q$-multinominal coefficient and 
Tsallis $q$-entropy \cite{Tsallis88,Tsallis98,book}, i.e., 
the one-parameter ($q$) generalization
of Eq. \eqref{entropy}.

On the other hand, Kaniadakis
\cite{Kaniadakis-Scarfone,Kaniadakis02,Kaniadakis05} has proposed the
$\kappa$-generalized statistical mechanics based on the
$\kappa$-deformed function, which is another type of one-parameter
deformations for the exponential and logarithmic functions. Based
on the $\kappa$-deformed functions and the associated product 
operation ($\kappa$-product) we have already shown the
$\kappa$-generalization of Gauss' law of error \cite{Wada-Suyari}.

In this work we show the $\kappa$-generalizations of the Stirling approximation
and the relation between the
$\kappa$-multinominal coefficient and $\kappa$-entropy.
In the next section the $\kappa$-factorial is introduced
based on the $\kappa$-product. Then the $\kappa$-generalization
of the Stirling approximation is obtained. 
We see that the naive approach fails in order to relate 
the $\kappa$-multinominal coefficients with the $\kappa$-entropy.
In order to overcome this difficulty, we introduce a new kind of 
$\kappa$-product
in section 3 and show the explicit relation between the $\kappa$-multinominal
coefficients and the $\kappa$-entropy. 
The final section is devoted to our conclusions.

\section{$\kappa$-Stirling approximation}
Let us begin with the brief review of the $\kappa$-factorial
and its Stirling approximation.
The $\kappa$-generalized
statistics \cite{Kaniadakis-Scarfone,Kaniadakis02,Kaniadakis05}
is based on the $\kappa$-entropy
\begin{equation}
  S_{\kappa} \equiv -\sum_i p_i \kln (p_i),
\end{equation}
where $\kappa$ is a real parameter in the range $(-1, 1)$, and
the $\kappa$-logarithmic function is
defined by
\begin{align}
 \kln (x) &\equiv \frac{x^{\kappa}-x^{-\kappa}}{2 \kappa}.
\end{align}
The inverse function is the $\kappa$-exponential function defined by
\begin{align}
 \kexp (x) &\equiv \left[
         \sqrt{1+\kappa^2 x^2}+\kappa x \right]^{\frac{1}{\kappa}}.
\end{align}
In the limit of $\kappa \to 0$ the both $\kappa$-logarithmic
and $\kappa$-exponential functions reduce to
the standard logarithmic and exponential functions, respectively.
We thus see that $S_{\kappa}$ reduces to 
the BGS entropy of Eq. \eqref{BGS} in the limit of $\kappa \to 0$.

Based on the above $\kappa$-deformed functions,
the $\kappa$-product is defined by
\begin{align}
   x \kprod y
   &\equiv \kexp \left[\kln (x) + \kln (y) \right] \nonumber \\
   &= \left[
      \left( \frac{ x^{\kappa}-x^{-\kappa} }{2} \right)+
      \left( \frac{ y^{\kappa}-y^{-\kappa} }{2} \right)+
      \sqrt{
      1+\left\{
       \left(\frac{ x^{\kappa}-x^{-\kappa}}{2}\right)+
        \left(\frac{ y^{\kappa}-y^{-\kappa}}{2}\right)
     \right\}^2
     } \; \;
     \right]^{\frac{1}{\kappa}},
   \label{kappa-product}
\end{align}
which reduces to the standard product $x \cdot y$ in the limit of
$\kappa \to 0$.
Similarly the $\kappa$-division is defined by
\begin{align}
   x \kdiv y
   &\equiv \kexp \left[\kln (x) - \kln (y) \right] \nonumber \\
   &= \left[
      \left( \frac{ x^{\kappa}-x^{-\kappa} }{2} \right)-
      \left( \frac{ y^{\kappa}-y^{-\kappa} }{2} \right)+
      \sqrt{
      1+\left\{
       \left(\frac{ x^{\kappa}-x^{-\kappa}}{2}\right)-
        \left(\frac{ y^{\kappa}-y^{-\kappa}}{2}\right)
     \right\}^2
     } \; \;
     \right]^{\frac{1}{\kappa}},
\end{align}
which reduces to the standard division $x / y$ in the limit of
$\kappa \to 0$.
By utilizing this $\kappa$-product,
the $\kappa$-factorial $n!_{\kappa}$ with $n \in \mathbb{N}$ is
defined by
\begin{align}
   n!_{\kappa} &\equiv 1 \kprod 2 \kprod \cdots \kprod n \nonumber \\
  & = \left[ \sum_{k=1}^{n}
      \left( \frac{k^{\kappa}- k^{-\kappa}}{2} \right)
       + \sqrt{
          \left\{ \sum_{k=1}^{n}
      \left(\frac{k^{\kappa}- k^{-\kappa}}{2} \right)   \right\}^2 +1}
    \; \right]^{\frac{1}{\kappa}}
 = \kexp \left[ \sum_{k=1}^{n} \kln (k) \right].
\end{align}

Now we come to the Stirling approximation of the $\kappa$-factorials. 
For sufficient large $n$, the summation is well approximated with the
integral as follows.
\begin{align}
   \kln( n!_{\kappa}) &=\sum_{k=1}^{n} \kln (k)
  \approx \int_{0}^{n} dx \kln (x)
 = \frac{n^{1+\kappa}}{2\kappa (1+\kappa)}
     - \frac{n^{1-\kappa}}{2\kappa (1-\kappa)}.
 \label{k-stirling}
 \end{align}
Clearly this reduces to the standard Stirling approximation as
\begin{align}
   \lim_{ \kappa \to 0 } \kln( n!_{\kappa} )
  \approx \lim_{ \kappa \to 0 } \frac{n}{1-\kappa^2} \left\{
   \kln(n) - \frac{n^{\kappa}+n^{-\kappa}}{2} \right\}
  = n \left( \ln n - 1 \right).
\end{align}

Next the $\kappa$-multinominal coefficient is defined by utilizing
the $\kappa$-product and $\kappa$-division as follows
\begin{align}
   \begin{bmatrix} n \\ n_1 \cdots  n_k
   \end{bmatrix}_{\kappa}
   \equiv n!_{\kappa} \kdiv
        \left( n_1!_{\kappa} \kprod \cdots \kprod n_k !_{\kappa} \right),
  \label{k-multi}
\end{align}
where we assume
\begin{align}
   n = \sum_{i=1}^k n_i.
\end{align}
In the limit of $\kappa \to 0$, Eq. \eqref{k-multi} reduces to the
standard multinominal coefficient of Eq. \eqref{multi}.

Let us try to relate the $\kappa$-multinominal coefficients
with the $\kappa$-entropy.
Taking the $\kappa$-logarithm of Eq. \eqref{k-multi} and
applying the $\kappa$-Stirling approximation leads to 
\begin{align}
  \kln \begin{bmatrix} n \\ n_1 \cdots  n_k
   \end{bmatrix}_{\kappa}
  &= \kln \left( n!_{\kappa} \right)
       - \sum_{i=1}^{k} \kln \left( n_i!_{\kappa} \right),
\end{align}
we then obtain
\begin{align}
\kln \begin{bmatrix} n \\ n_{1} \cdots  n_{k}
   \end{bmatrix}_{\kappa}
  \; \approx
  \frac{n^{1+\kappa}}{2 \kappa (1+\kappa)}
       \left\{1-\sum_{i=1}^{k} \left(\frac{n_i}{n}\right)^{1+\kappa}\right\}
 + \; \frac{n^{1-\kappa}}{2 \kappa (1-\kappa)}
       \left\{ \sum_{i=1}^k \left(\frac{n_i}{n}\right)^{1-\kappa} -1
       \right\}.
  \label{kln-mul}
\end{align}
From this relation, we see that the above naive approach fails.
Since the r.h.s. consists of the two terms with different factors (one is
proportional to $n^{1+\kappa}/(1+\kappa)$ and the other is
proportional to $n^{1-\kappa}/(1-\kappa)$), this cannot be
proportional to the $\kappa$-entropy, which can be written by
\begin{align}
 S_{\kappa} \left(\frac{n_1}{n}, \ldots, \frac{n_k}{n}\right)
 &\equiv
  -\sum_{i=1}^{k} \left(\frac{n_i}{n}\right) \kln \left(\frac{n_i}{n}\right)
\nonumber \\
 & = \frac{1}{2\kappa}
       \left\{
 1-\sum_{i=1}^{k} \left(\frac{n_i}{n}\right)^{1+\kappa} \right\} +
\frac{1}{2\kappa}
       \left\{
 \sum_{i=1}^k \left(\frac{n_i}{n}\right)^{1-\kappa}-1
\right\}.
\end{align}

\section{Introducing a new $\kappa$-product}
In order to overcome the above difficulty, we introduce another
$\kappa$-generalization of the products based on the following
function defined by
\begin{align}
 \ku (x) \equiv \frac{x^{\kappa}+x^{-\kappa}}{2}
= \cosh \big( \kappa \ln(x) \big).
\end{align}
We here call it the $\kappa$-generalized unit function since
$\lim_{\kappa \to 0} \ku (x) = 1$.
The basic properties of $\ku(x)$ are as follows.
\begin{align}
 \ku (x) &= \ku{ \left( \frac{1}{x} \right)}, \\
 \ku (x) &\ge 1, \; \textrm{because} \;
  \ku(x)-1=\frac{(x^{\frac{\kappa}{2}}-x^{-\frac{\kappa}{2}})^2}{2} \ge 0.
\end{align}
The inverse function of $\ku(x)$ can be defined by
\begin{equation}
 \ku^{-1} (x) \equiv \left[
         \sqrt{x^2-1}+ x \right]^{\frac{1}{\kappa}},
   \qquad (x \ge 1).
  \label{inv_ku}
\end{equation}
In Ref. \cite{Z_k},  the canonical partition function associated with 
the $\kappa$-entropy is obtained in terms of this $\ku$ function.
Note that
\begin{align}
  \kln (x) =
   \frac{x^{\kappa}-x^{-\kappa}}{2 \kappa}
  = \frac{1}{\kappa} \; \sinh \big( \kappa \ln(x) \big),
\end{align}
the two kinds of the $\kappa$-deformed functions 
($\kln(x)$ and $\kexp(x)$; $\ku(x)$ and $\ku^{-1}(x)$) are thus associated
each other. This can be seen from the following relations
\begin{align}
 \sqrt{1+\kappa^2 \kln^2(x)} &= \ku (x), \\
 \ku^{-1} \left(\sqrt{1+\kappa^2 x^2} \right) &= \kexp (x),
\quad \textrm{for} \; x\ge 0.
\end{align}

Now by utilizing these functions, a new $\kappa$-product is defined by
\begin{align}
   x \uprod y
   &\equiv \ku^{-1} \left[\ku (x) + \ku (y) \right] \nonumber \\
   &= \left[
      \left( \frac{ x^{\kappa}+x^{-\kappa} }{2} \right)+
      \left( \frac{ y^{\kappa}+y^{-\kappa} }{2} \right)+
      \sqrt{
      \left( \frac{ x^{\kappa}+x^{-\kappa} }{2}
             + \frac{ y^{\kappa}+y^{-\kappa} }{2}  \right)^2 -1
     } \; \;
     \right]^{\frac{1}{\kappa}}
   \label{u-product}
\end{align}
Similarly the corresponding $\kappa$-division is defined by
\begin{align}
   x \udiv y
   &\equiv \ku^{-1} \left[\ku (x) - \ku (y) \right] \nonumber \\
   &= \left[
      \left( \frac{ x^{\kappa}+x^{-\kappa} }{2} \right)-
      \left( \frac{ y^{\kappa}+y^{-\kappa} }{2} \right)+
      \sqrt{
      \left( \frac{ x^{\kappa}+x^{-\kappa} }{2}
             - \frac{ y^{\kappa}+y^{-\kappa} }{2}  \right)^2 -1
     } \; \;
     \right]^{\frac{1}{\kappa}},
   \label{u-division}
\end{align}
where $\left(\frac{ x^{\kappa}+x^{-\kappa} }{2}\right) - 
\left(\frac{y^{\kappa}+y^{-\kappa} }{2} \right) \ge 1$.\\ 
Note that $\lim_{\kappa \to 0}
x \uprod y = \infty$, consequently there is no corresponding operation
in the standard case of $\kappa=0$. 
However unless $\kappa = 0$ this new product satisfies
\begin{align}
   x \uprod y &= y \uprod x, \quad &\textrm{comutativity} \\
   (x \uprod y) \uprod z &= x \uprod (y \uprod z). &\textrm{associativity}
\end{align}
There exists no unit element within a real number. Therefore real
numbers and this product consist a semigroup!\\
Note also that $x \udiv x \ne 1$ because $\ku^{-1}(0)$ does not exist
by the definition Eq. \eqref{inv_ku}. However we see that
the following identity holds.
\begin{align}
   x \uprod y \udiv x = y, \quad (\kappa \ne 0).
  \label{k-identity}
\end{align}

Using the new $\kappa$-product, the associated $\kappa$-factorial can be 
introduced as
\begin{align}
   n!^{\kappa} &
\equiv 1 \uprod 2 \uprod \cdots \uprod n \nonumber \\
  & = \left[ \sum_{k=1}^{n}
      \left( \frac{k^{\kappa}+ k^{-\kappa}}{2} \right)
       + \sqrt{
          \left\{ \sum_{k=1}^{n}
      \left( \frac{k^{\kappa}+k^{-\kappa}}{2} \right)   \right\}^2 -1}
    \; \right]^{\frac{1}{\kappa}}.
\end{align}
Similar to Eq. \eqref{k-stirling}, 
the $\kappa$-Stirling approximation can be obtained as 
\begin{align}
   \ku( n!^{\kappa}) &=\sum_{k=1}^{n} \ku (k)
\approx \int_{0}^{n} dx \; \ku (x)  \nonumber \\
 &= \frac{n^{1+\kappa}}{2 (1+\kappa)}
     + \frac{n^{1-\kappa}}{2 (1-\kappa)}.
   \label{u-stirling}
\end{align}
Also the corresponding $\kappa$-multinominal coefficient can be defined by
\begin{align}
   \begin{bmatrix} n \\ n_1 \cdots  n_k
   \end{bmatrix}^{\kappa}
  &\equiv n!^{\kappa} \udiv
        \left( n_1!^{\kappa} \uprod \cdots \uprod n_k !^{\kappa} \right).
  \label{uk-multi}
\end{align}
Applying the above Stirling approximation to Eq. \eqref{uk-multi} we obtain
\begin{align}
\ku \begin{bmatrix} n \\ n_{1} \cdots  n_{k}
   \end{bmatrix}^{\kappa}
  \approx &
  \frac{n^{1+\kappa}}{2 (1+\kappa)}
       \left\{1-\sum_{i=1}^{k} \left(\frac{n_i}{n}\right)^{1+\kappa}\right\}
 - \; \frac{n^{1-\kappa}}{2(1-\kappa)}
       \left\{ \sum_{i=1}^k \left(\frac{n_i}{n}\right)^{1-\kappa} -1 \right\}.
 \label{ku-mul}
\end{align}
This is the complemental relation to Eq. \eqref{kln-mul},
and by combining Eqs \eqref{kln-mul} and \eqref{ku-mul} we have
\begin{align}
\kln \begin{bmatrix} n \\ n_{1} \cdots  n_{k}
   \end{bmatrix}_{\kappa}& \pm
  \frac{1}{\kappa} \; \ku \; \begin{bmatrix} n \\ n_{1} \cdots  n_{k}
   \end{bmatrix}^{\kappa}
  \approx \pm \;
  \frac{n^{1 \pm \kappa}}{\kappa(1 \pm \kappa)}
     \left\{1-\sum_{i=1}^{k} \left(\frac{n_i}{n}\right)^{1 \pm \kappa}\right\},
\end{align}
We thus obtain the final result
\begin{align}
   &\left( \frac{1-\kappa} {2 n^{1-\kappa}} \right)
   \left( \kln \begin{bmatrix} n \\ n_{1} \cdots  n_{k}
   \end{bmatrix}_{\kappa} -
  \frac{1}{\kappa} \ku \; \begin{bmatrix} n \\ n_{1} \cdots  n_{k}
   \end{bmatrix}^{\kappa}
   \right) \nonumber \\
   & \qquad \qquad \qquad +
   \left( \frac{1+\kappa} {2 n^{1+\kappa}} \right)
   \left( \kln \begin{bmatrix} n \\ n_{1} \cdots  n_{k}
   \end{bmatrix}_{\kappa} +
  \frac{1}{\kappa} \ku \; \begin{bmatrix} n \\ n_{1} \cdots  n_{k}
   \end{bmatrix}^{\kappa}
   \right) \nonumber \\
  &= 
  \frac{1}{2\kappa}
       \left\{1-\sum_{i=1}^{k} \left(\frac{n_i}{n}\right)^{1+\kappa}\right\}
  +  \frac{1}{2 \kappa}
       \left\{ \sum_{i=1}^k \left(\frac{n_i}{n}\right)^{1-\kappa} -1 \right\}
= S_{\kappa}\left(\frac{n_1}{n}, \ldots, \frac{n_k}{n}\right).
 \label{kmul-Sk}
\end{align}

Note that since
\begin{equation}
\lim_{\kappa \to 0}
  \frac{1}{\kappa} \; \ku \; \begin{bmatrix} n \\ n_{1} \cdots  n_{k}
   \end{bmatrix}^{\kappa} = 0,
  \label{lim_u_mulcoeff}
\end{equation}
as shown in the Appendix,
Eq. \eqref{kmul-Sk} reduces to the standard case Eq. \eqref{entropy}
in the limit of $\kappa \to 0$.

\section{Conclusion}
We have generalized Stirling approximation of the factorials and 
multinominal coefficients
based on the $\kappa$-deformed function introduced by Kaniadakis, 
which is different one-parameter generalization 
from the $q$-deformed functions by Tsallis.
In order to relate the $\kappa$-generalized
multinominal coefficients to the $\kappa$-entropy, we
showed the naive approach, which is similar to that \cite{Suyari-b} for 
the $q$-generalization, failed.
In order to overcome this difficulty, we have introduced a new kind of 
the $\kappa$-product operations, which never exist in the standard
case of $\kappa=0$, and have obtained the relation between 
the $\kappa$-generalized multinominal coefficients and the $\kappa$-entropy.

The final result Eq. \eqref{kmul-Sk} clearly states that the maximizing 
$\kappa$-entropy is equivalent to the maximizing the l.h.s. 
of Eq. \eqref{kmul-Sk} as same as
in the standard case of Eq. \eqref{entropy}.  
A next step in a future work is thus to clarify 
what microscopic states are described by the both $\kappa$-multinominal 
coefficients Eqs. \eqref{k-multi} and \eqref{uk-multi}.


\section{Appendix}
We here show the proof of Eq. \eqref{lim_u_mulcoeff}.
\begin{equation}
  \frac{1}{\kappa} \; \ku \; \begin{bmatrix} n \\ n_{1} \cdots  n_{k}
   \end{bmatrix}^{\kappa} =
\frac{1}{\kappa} \left( \sum_{\ell=1}^{n} \ku(\ell) - \sum_{i=1}^k
\sum_{j=1}^{n_i} \ku(j) \right)
 ,
\end{equation}
In the limit of $\kappa \to 0$ the numerator reduces to $n - \sum_{i=1}^k n_i$
since $\lim_{\kappa \to 0} \ku = 1$.
Consequently both the denominator $\kappa$ and the numerator become null.
Then applying l'Hopital's rule and using the relation 
\begin{equation}
 \frac{d}{d \kappa} \; \ku(x) \; 
   = \ln x \left( \frac{x^{\kappa}-x^{-\kappa}}{2} \right),
\end{equation}
we obtain Eq. \eqref{lim_u_mulcoeff}. 



\begin{thebibliography}{99}
\bibitem{Suyari05}
H.~Suyari and M.~Tsukada,
IEEE Trans. Inform. Theory, \textbf{51} (2005) 753-757.
%
\bibitem{Suyari-a}
H.~Suyari, ``q-Stirling's formula in Tsallis statistics'',
arXiv: cond-mat/0401541.
%
\bibitem{Suyari-b}
H.~Suyari,
``Mathematical structure derived from the q-multinomial coefficient
in Tsallis statistics'', arXiv: cond-mat/0401546.

\bibitem{Wang}
L. Nivanen, A. Le Mehaute, and Q. A. Wang,                         
Rep. Math. Phys., \textbf{52} (2003) 437-444.

\bibitem{Borges}
E. P. Borges,                                                                  
Physica A, \textbf{340} (2004) 95-101.             

\bibitem{Tsallis88}
C. Tsallis, Possible generalization of Boltzmann-Gibbs
statistics, J. Stat. Phys. \textbf{52} (1988) 479-487.

\bibitem{Tsallis98}
C. Tsallis, R.S. Mendes, A.R. Plastino,
Physica A \textbf{261} (1998) 534-554.

\bibitem{book}
M.~Gell-Mann, C.~Tsallis, \textit{Nonextensive Entropy:
Interdisciplinary Applications} (Oxford University Press,
Oxford 2004).


%

\bibitem{Kaniadakis-Scarfone}
G.~Kaniadakis and A.M. Scarfone, Physica A, \textbf{305} (2002) 69.

\bibitem{Kaniadakis02}
G.~Kaniadakis, Phys. Rev. E \textbf{66} (2002) 056125.

\bibitem{Kaniadakis05}
G. Kaniadakis,
 Phys. Rev. E \textbf{72} (2005) 036108.

\bibitem{Wada-Suyari}
T.~Wada and H.~Suyari, ``$\kappa$-generalization of Gauss' law of error'',
Phys. Lett. A \textbf{} (2005) in press.

\bibitem{Z_k}
A.M.~Scarfone and T.~Wada, ``Canonical partition function for
anomalous systems described by the $\kappa$-entropy'',
arXiv: cond-mat/0509364


\end{thebibliography}
\end{document}